\documentclass[12pt]{article}
\usepackage{osajnl2}
\usepackage{amssymb}   
\usepackage{amsmath}

\begin{document}

\title{Programmable unitary spatial modes manipulation}
\date{\today}
\author{Jean-Fran\c cois Morizur,$^{1,2,*}$ Lachlan Nicholls,$^{1}$ Pu Jian,$^{1,2}$ Seiji Armstrong,$^{1}$ Nicolas Treps,$^{2}$ Boris Hage,$^{1}$ Magnus Hsu,$^{3}$ Warwick Bowen,$^{3}$ Jiri Janousek,$^{1}$ Hans-A. Bachor$^{1}$}
\address{$^{1}$ARC Centre of Excellence for Quantum-Atom Optics, The Australian National University, Canberra, Australia}
\address{$^{2}$Laboratoire Kastler Brossel, UniversitŽ Pierre et Marie Curie Ð Paris 6, ENS, CNRS, Paris, France}
\address{$^{3}$University of Queensland, School of Mathematics and Physics, Brisbane, Australia}
\address{$^{*}$ Corresponding author: morizur@spectro.jussieu.fr}

\maketitle 

\begin{abstract}
Free space propagation and conventional optical systems such as lenses and mirrors all perform spatial unitary transforms. However, the subset of transforms available through these conventional systems is limited in scope.\\
We present here a unitary programmable mode converter (UPMC) capable of performing any spatial unitary transform of the light field. It is based on a succession of reflections on programmable deformable mirrors and free space propagation. We first show theoretically that a UPMC without limitations on resources can perform perfectly any transform. We then build an experimental implementation of the UPMC and show that, even when limited to three reflections on an array of 12 pixels, the UPMC is capable of performing single mode tranforms with an efficiency greater than $80\%$ for the first 4 modes of the TEM basis.  \\
\end{abstract}


\section{Introduction}

Optical systems manipulate light by transforming an input light field into a desired output. Unitary optical systems do not add nor substract energy to the light in question; they change its spatial characteristics. Put another way, The light is reshaped without any loss of the information carried by its photons \cite{Fade:2008p282}.
Ranging in scale from the large (telescope) \cite{Preumont:2009p1620} to the small (microscope, CD) \cite{Torok:2007p1916}, ranging in complexity from the simplicity of a single lens to the resource heavy nature of adaptive optics in ophtalmoloscopy \cite{Roorda:2002p1991}, these systems perform a specific unitary transform on the input light field. Conventionally, a specific set of optical components are selected and assembled to perform a specific unitary transform. For example, to rotate a beam along its propagation axis one may use two dove prisms \cite{Leach:2002p2000}, or to rescale a field profile a telescope may be used \cite{Herschel:1861p2062}.\\ 
In this paper, we present a unitary programmable mode converter (UPMC). This device can be programmed to perform any desired unitary transform. At the heart of the UPMC are deformable mirrors whose topographies are controlled by actuators \cite{Potsaid:2008p2108}. A succession of reflections upon the deformable mirrors are performed, while the reflections are separated by free space propagation and a lens which perform a Fourier Transform (FT) of the field's spatial profile. \\
The structure of the paper is as follows. We begin with a theoretical demonstration of the statement that any desired unitary transform of the light field can be achieved with such a succession of reflections, provided that the topographies of the deformable mirrors are adequately selected. This mathematical demonstration \cite{Borevich:1981p758} provides a tractable yet inefficient solution consisting of a succession of topographies, involving a large number of reflections. Since the cost and complexity scale with the number of reflections on deformable mirrors we introduce a measure of how well an optical system approaches the desired transform using limited resources.\\
We then present and characterize an experimental UPMC with a limited number of reflections. An optimization algorithm is introduced to find the best topographies for the deformable mirrors, and we compare the experimental performance of the UPMC to propagation simulation results. In this proof-of-principle experiment, we only characterize the UPMC performances in transforming a single specified input field into a specified output. The propagation simulation, whose validity has been verified against experimental results for these single mode transforms, is then extended to compute the performances of the UPMC for general multimode manipulations.

\section{Theoretical Considerations}

Conventionally, changing the spatial profile of a beam of light is a destructive process: through phase control and attenuation (local attenuation \cite{Hsieh:2007p2173}, spatial filtering \cite{Bagnoud:2004p2111}...), a given profile can be transformed into any other profile, albeit with smaller total intensity. Here, we demonstrate mathematically that changing arbitrarily the spatial profile of a beam need not be destructive. More generally, we show that any unitary optical transform is achievable. \\
To begin with, let us define the mathematical framework for the linear optical processes arising in the UPMC, arising from the deformable mirrors, and including FTs.
Without loss of generality, we consider a beam of monochromatic, linearly polarized light of wavelength $\lambda$ propagating along the $\mathbf{z}$ axis. The spatial distribution of magnitude and phase of the beam in the plane $(\mathbf{x},\mathbf{y},z=0)$ is the transverse profile of the beam: $E\left(x,y\right)=A\left(x,y\right)e^{i\phi\left(x,y\right)}$. This profile can be decomposed in a transverse mode basis such as the Transverse ElectroMagnetic (TEM) modes:
\begin{equation}
E(x,y)=\sum_{m\in \mathbb N, n \in \mathbb N} a_{mn} TEM_{mn}(x,y)
\end{equation}
The $a_{mn}$ are the complex coefficients of the decomposition of $E(x,y)$ in the TEM basis.\\ 
An optical system transforms an input field $I(x,y)$ into an output $O(x,y)$, with different inputs producing different outputs. Any linear optical system is fully characterized by its action in a transverse basis: the output of each mode of the basis through the optical system can be decomposed in the same basis, thus providing a matrix description of the transform. For example, the transform defined by 
\begin{equation}
U_{T}=
\begin{pmatrix}
\frac{1}{\sqrt{2}} & \frac{1}{\sqrt{2}} & 0 & 0 & ...\\
\frac{1}{\sqrt{2}} & -\frac{1}{\sqrt{2}} & 0 & 0 & ...\\
0 & 0 & 1 & 0 & ...\\
0 & 0 & 0 & 1 & ... \\
... & ... & ... & ... & ... \\
\end{pmatrix}
\end{equation}
in the TEM basis acts as a two mode beamsplitter wherein the spatial modes $TEM_{00}$ and $TEM_{10}$ are mixed together, while all the other modes in $TEM_{mn}$ remain unaffected.\\
In the specific case of a unitary optical system, the total intensity of the output is identical to the input's. This entails that the matrix describing the transform is itself unitary. A unitary matrix can also be seen as a basis change; it transforms an input basis into another output basis. Any combination of lenses, mirrors and free space propagation is unitary, but these Gaussian elements \cite{ESIEGMAN:1986p578} give access to only a small subset of all the unitary transforms \cite{ESIEGMAN:1986p578}. For example, the $TEM_{mn}$ modes are eigenmodes of the propagation through these elements; their size changes, but their intensity distribution does not. To that extent, these Gaussian elements are not sufficient to transform a $TEM_{00}$ mode into a $TEM_{10}$ mode. \\

The UPMC aims at performing any unitary transform, including transforms that change a $TEM_{00}$ input into a $TEM_{10}$ output. This means that non Gaussian optical elements are needed, such as deformable mirrors. A programmable deformable mirror is a surface whose topography $z(x,y)$ can be defined by the user. When the beam hits the mirror, the field profile is transformed into
\begin{equation}
E(x,y) \rightarrow e^{i\phi_{DM}(x,y)}E(x,y)
\end{equation}
with $\phi_{DM}(x,y)=2\pi \frac{z(x,y)}{\lambda}$. While the decomposition of the transverse profile $E(x,y)$ on the TEM basis remains constant throughout the propagation of the beam, the reflection on a deformable mirror changes the coefficients of the decomposition. In order to describe the unitary transform induced by the reflection, we introduce a different transverse basis. \\
Due to the finite number of actuators deforming the mirrors, we consider the discretization of the transverse profile on a pixel basis: $E_{ij}=E(x_i,y_j)$. $x_i$ and $y_j$ are the $(x,y)$ coordinates of pixel $(i,j)$ with $i \in \mathbb Z$ and $j \in \mathbb Z$. When the size of the pixel $\Delta_{pix}$ is small compared to the variations of the transverse field $E(x,y)$ along the $\mathbf{x}$ and $\mathbf{y}$ axis, the transverse profile $E(x,y)$ is adequately described by its discretization $E_{ij}$. Moreover, since the transverse extension of a physical beam is finite, $E(x,y)$ is adequately described by $E_{ij}$ with $\vert i \vert \le N_{pix}/2 $ and $\vert j \vert  \le N_{pix}/2$ for a large enough $N_{pix}$.\\
In this basis, the reflection on a deformable mirror is the transform $E_{ij} \rightarrow e^{i\phi_{ij}}E_{ij}$ with $e^{i\phi_{ij}} = e^{i\phi_{DM}(x_i,y_j)}$.
The table of $E_{ij}$ can be reorganized row by row into a single vector $\mathbf{E}_k=E_{ij}$ with $k \in A$ ($A=\{1,2, ... , n=N_{pix}^2\}$). The deformable mirror transforms becomes $\mathbf{E} \rightarrow U_{DM} \left(\mathbf{\phi}\right) \mathbf{E}$ with 
\begin{equation}
U_{DM}\left(\mathbf{\phi}\right)=
\begin{pmatrix}
e^{i\phi_1} & 0 & 0 & ... & 0\\
0 & e^{i\phi_2} & 0 & ... & 0\\
0 & 0 & e^{i\phi_3} & ... & 0\\
... & ... & ... & ... & ...\\
0&0&0& ... & e^{i\phi_n}
\end{pmatrix}
\end{equation}
In the pixel basis, the set of all possible $U_{DM}$ forms the subgroup $D\mathbb U$ of the unitary group $\mathbb U$.\\

Any unitary transform in the group $D\mathbb U$ is a local phase manipulation and does not change the intensity distribution of the beam. In order to transform a field input into any other output, the intensity distribution also needs to be changed. Propagation through gaussian elements do not change the intensity distribution of $TEM_{mn}$ modes. However, for the modes which are decomposed in multiple $TEM_{mn}$ modes, the Gouy phase shift induced by propagation changes the intensity distribution \cite{ESIEGMAN:1986p578}.\\
Let us consider more specifically the unitary transform performed by the combination of a lens of focal length $f_0$ and free-space propagation before and after the lens of a distance $f_0$. These elements perform a FT on the spatial profile of the beam, and the profile is rescaled: for an input beam with transverse size parameter $\omega_{in}$, the typical transverse size of the output is $\omega_{out}=2\pi\lambda f_0 / \omega_{in}$.\\
Let us name the unitary matrix of this FT in the pixel basis $U_{FT}$;  $\mathbf{E} \rightarrow U_{FT}\mathbf{E}$.
We choose the pixel size $\Delta_{pix}$ small enough and the number of pixels $N_{pix}$ large enough to make sure there are no zero elements in $U_{FT}$. Typically, $\Delta_{pix}^2N_{pix}=2\pi\lambda f_0$.\\

We have thus far provided linear algebra models for the two components of the UPMC; the reflections on deformable mirrors and the FTs. We now consider a succession of these components, and using group theory, we show that they can provide any unitary transform. Let us name $\mathbb H$ the set of all possible optical transforms provided by such a succession. $\mathbb H$ is a subgroup of $\mathbb U$ that contains $D\mathbb U$ and $U_{FT}$. We want to show that $\mathbb H$ is $\mathbb U$.\\
Let us now consider $\mathbb U_{ij}$ the subgroup of $\mathbb U$ that contains all the matrices of the form 
\begin{equation}
T_{ij}(\theta)=
\begin{pmatrix}
1 & 0 & ... & 0 & ... & 0 & ... & 0\\
0 & 1 & ... & 0 & ... & 0 & ... & 0\\
... & ... & ... & ... & ... & ... & ... & ...\\
0 & 0 & ... & cos(\theta) & ... & sin(\theta) & ... & 0\\
... & ... & ... & ... & 1 & ... & ... & ...\\
0 & 0 & ... & -sin(\theta) & ... & cos(\theta) & ... & 0\\
... & ... & ... & ... & ... & ... & ... & ...\\
0 & 0 & ... & 0 & ... & 0 & ... & 0\\
0 & 0 & ... & 0 & ... & 0 & ... & 1\\
\end{pmatrix}
\end{equation}
where the $sin(\theta)$  and $cos(\theta)$ terms are in the ith row and jth column. Using a reorganisation of the matrix coefficients that can be found in \cite{Borevich:1981p758}, it is possible to build $T_{ij}(\theta)$ for any triplet $(i,j,\theta)$ with a succession of $U_{FT}$ and $U_{DM}(\mathbf{\phi})$, as long as there is no zero element in $U_{FT}$. This means that all the $T_{ij}(\theta)$ are in $\mathbb H$, so all the $\mathbb U_{ij}$ are in $\mathbb H$.
We now know that $\mathbb H$ is a subgroup of the unitary matrices $\mathbb U$ that contains all the rotations $T_{ij}(\theta)$ and the diagonal matrices $U_{DM}(\mathbf{\phi})$. It is easy to show (see for example \cite{Serre:2002p2423}) that with a succession of  $T_{ij}(\theta)$ and $U_{DM}(\mathbf{\phi})$ any unitary matrix can be built. Since $\mathbb H$ is a group, $\mathbb H$ contains all these successions. This means that $\mathbb H$ is $\mathbb U$: $\mathbb H$, the set of optical transforms formed by all the successions of reflections on deformable mirrors and spatial FTs, encompasses all the unitary transforms. Any desired unitary transform has a systematic decomposition in terms of reflections on specific topographies and FTs. \\ 
A finite sequence of $U_{DM}$ and $U_{FT}$ sufficient to build any $T_{ij}(\theta)$ is presented in \cite{Borevich:1981p758}. Such a systematic construction requires 17 reflections on deformable mirrors, separated by FTs.
To build any unitary matrix only a finite number of $T_{ij}(\theta)$ are required. Consequently a finite number of reflections on deformable mirrors separated by FTs is sufficient to build any unitary matrix. When the number of pixels is increased, the number of $T_{ij}(\theta)$ required to build a general unitary matrix increases. \\
As a conclusion, this theoretical study showed that any kind of unitary transform can be performed using a finite succession of reflections on deformable mirrors and FTs. However, experimentally the number of reflections on deformable mirrors can be limited. We need to introduce a measure to evaluate how well an optical system built with limited resources performs a desired transform.\\

Let us consider that the desired unitary transform D is defined by the $n$ orthonormal output field modes $O_i(x,y)$ for the orthonormal input modes $I_i(x,y)$ (with $i\in K$, $K=\{1,2,...,n\}$). For example D can be defined on a single mode. In this case we want a specific output for a given input, but the action of D on all the other inputs is irrelevant. Or D can be defined on some or all the modes of a transverse basis. In the latter case, the desired transform matrix is completely specified.\\
For the same inputs $I_i(x,y)$ an optical system A has the outputs $O'_i(x,y)$. When D is defined on a single mode, a standard measure of how well A performs the transform D is the intensity overlap between the mode  $O_1(x,y)$ and the mode  $O'_1(x,y)$. We introduce the coefficient $\alpha$, which is a generalization of this single mode case to all unitary transforms. It combines all the output overlaps, and is sensitive to the phase between the overlaps: 
\begin{equation}
\alpha=\left\vert \sum_{i\in K} \iint_{(\mathbf{x},\mathbf{y})} \bar{O}_i(x,y) O'_i(x,y) \right\vert 
\end{equation}
If we decompose $O_i(x,y)=\sum_{m\in \mathbb N, n \in \mathbb N} o_{i,m,n} TEM_{mn}(x,y)$ and $O'_i(x,y)=\sum_{m\in \mathbb N, n \in \mathbb N} o'_{i,m,n} TEM_{mn}(x,y)$, $\alpha$ can be written as
\begin{equation}
\alpha=\left\vert \sum_{i \in K} \sum_{m\in \mathbb N, n \in \mathbb N} \bar{o}_{i,m,n} o'_{i,m,n} \right\vert 
\end{equation}
In this last notation, $\alpha$ is a scalar product between the unitary matrices of D and A in the TEM basis.\\ 

When $n=1$, $\alpha^2$ is the mode conversion efficiency: it is the proportion of power effectively tranfered from the input mode into the desired output. For multimode transformations, i.e. higher values of $n$, we introduce the transform quality $\alpha_n^2=\frac{\alpha^2}{n^2}$, a generalization of the mode conversion efficiency. $\alpha_n^2$ is normalized to compare transforms with different input mode numbers on the same $0$ to $1$ scale. When $\alpha_n^2=1$, there is no difference between A and D, provided that we only consider the subspace formed by the input modes $I_i(x,y), i\in K$. The two transforms can still be different; they just differ outside of the considered input modes. $\alpha_n^2$ was chosen as the measure of the difference between the transforms because it gives the same importance to all the considered modes and coefficients. For a specific purpose, another measure of distance could be more appropriate. For example, when the relative phases between the output modes are irrelevant, the quantity $\beta=\sum_{i\in K} \left\vert \iint_{(\mathbf{x},\mathbf{y})} \bar{O}_i(x,y) O'_i(x,y) \right\vert$ is more appropriate.\\

For a given transform D, finding the optimal achievable transform A within the experimental constraints is an optimization problem. When a large number of reflections on deformable mirrors is possible, we can use the systematic decomposition sequence of D in $U_{FT}$ and $U_{DM}(\phi)$ presented above to perform the transform perfectly. However when the number of reflections is a constraint, there is no algebraic solution \cite{Wyrowski:1991p1437}. The problem then comes down to the optimization of a finite set of parameters (here the topographies of the deformable mirrors) and can be efficiently solved using a stochastic approach, using $\alpha_n^2$ as the optimization criterion. \\

In this section we showed that the UPMC has in theory the ability to perform any desired unitary transform. Since performing this transform perfectly is resource heavy, we introduce a measure $\alpha_n^2$, a generalization of the intensity overlap, to evaluate how well an optical system performs the desired transform.\\

\section{Experimental Demonstration}

An experimental implementation of the UPMC was built in order to verify the theoretical capabilities discussed in the previous section. 
We will first show that the experimental deformable mirror can be modelled reasonably well by the theoretical unitary transform $U_{DM}$. The optical set-up of the UPMC is then presented, designed to allow three reflections on the same deformable mirror. Finally, we characterize this experimental UPMC's ability to perform a range of single mode transforms. A propagation model of this UPMC, validated by these experimental results, can then be used to evaluate the performances of the UPMC to multimode transforms. \\

\begin{figure}[htbp]
\begin{center}
\mbox{\includegraphics[width=5.00in]{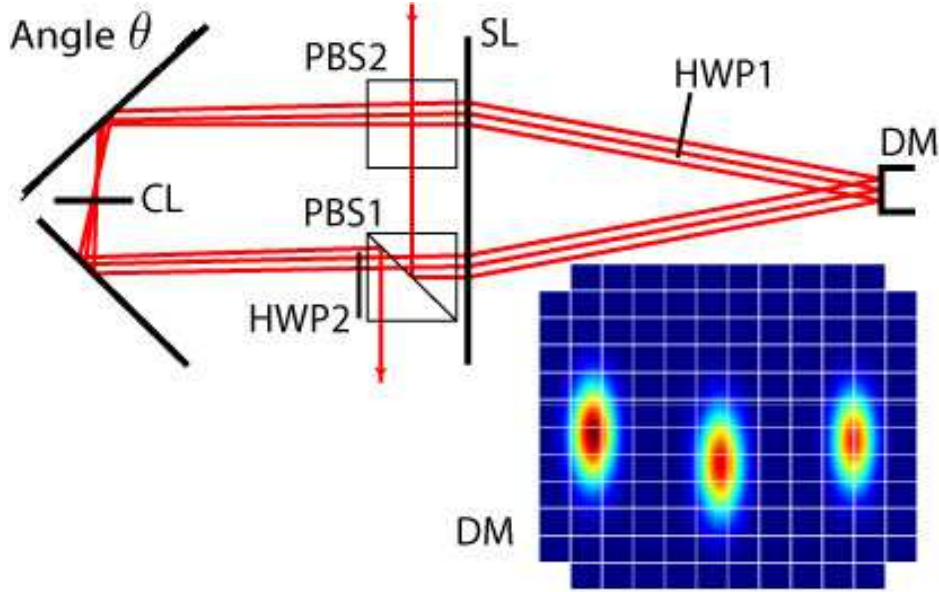}}
\caption{(Color online) The Beam is coupled into the UPMC by a reflection on the polarizing beamsplitter PBS1. It is then focused on to the deformable mirror DM. The beam first undergoes a 2D FT through the spherical lens SL, followed by a vertical FT going through the cylindrical lens CL, and finally another 2D FT going back through SL. The sliding half wave plates HWP1 and HWP2 are used to choose how many times the beam is reflected by the deformable mirror before being coupled out on PBS1. The deformable mirror representation in the inset shows the pixel layout and the measured positions and sizes of the beam for the 3 reflections, when the beam is a simple gaussian. Manipulation of the beam makes the spatial profile bigger than the simple gaussian, hence the small footprint of the simple gaussian compared to the size of the Deformable Mirror.}
\label{ScheUPMC}
\end{center}
\end{figure}

We use a Thorlabs multi-DM as a programmable deformable mirror. This device is a continuous membrane with a gold coating, controlled by 140 electrostatic actuators (laid out in a 12 by 12 square without the 4 corners). The schematic is displayed in the inset of Fig. \ref{ScheUPMC}. The actuators are computer controlled, with a settling time of $10$ms. There is no measurable modulation of the beam intensity or phase profile due to any flickering of the deformable mirror, as can be sometime found in the liquid crystal on silicon spatial light modulators (LCOS SLM) \cite{Tay:2009p2224}. \\
Comparing the deformable mirror with a highly reflective flat mirror, we can derive the optical power loss induced by the deformable mirror and its protective window. For each reflection we find a $4.2\% (\pm 0.5\%)$ loss compared to the flat mirror. Manufacturer specifications predict a $3.4\%$ loss; we can safely conclude that the only losses on the deformable mirror are due to the absorption on the gold coating and the scattering from the protective window. These two technical losses could be reduced by using a highly anti-reflective coated protective window and gold surfaces. In the approximation that the losses are (or can be made) negligible, it is justified to model the transform induced by a single reflection on the deformable mirror by the matrix $U_{DM}$ with the topography $z(x,y)$ controlled by the actuators. \\

For this proof of principle UPMC, we utilize a single deformable mirror. In order to satisfy the requirement of having multiple reflections on deformable surfaces, we choose to designate three separate areas of the deformable mirror, allowing for three successive independent reflections. A highly elliptic beam is used; the topography of the deformable mirror along the vertical axis is controlled by 12 actuators, while the horizontal axis remains flat (see inset in Fig. \ref{ScheUPMC}). Between each reflection a FT on the vertical axis is performed using a cylindrical lens and a spherical lens. We use half wave plates and polarizing beamsplitters to couple the beam in and out of the UPMC. A schematic of the general UPMC can be found in Fig. \ref{ScheUPMC}. This pratical implementation of the UPMC is limited in the maximum number of reflections (3) and in the number of actuators per reflection (12). To that extent, the ideal theoretical construction sequence presented above cannot be performed.\\

\begin{figure}[htbp]
\begin{center}
\mbox{\includegraphics[width=6.00in]{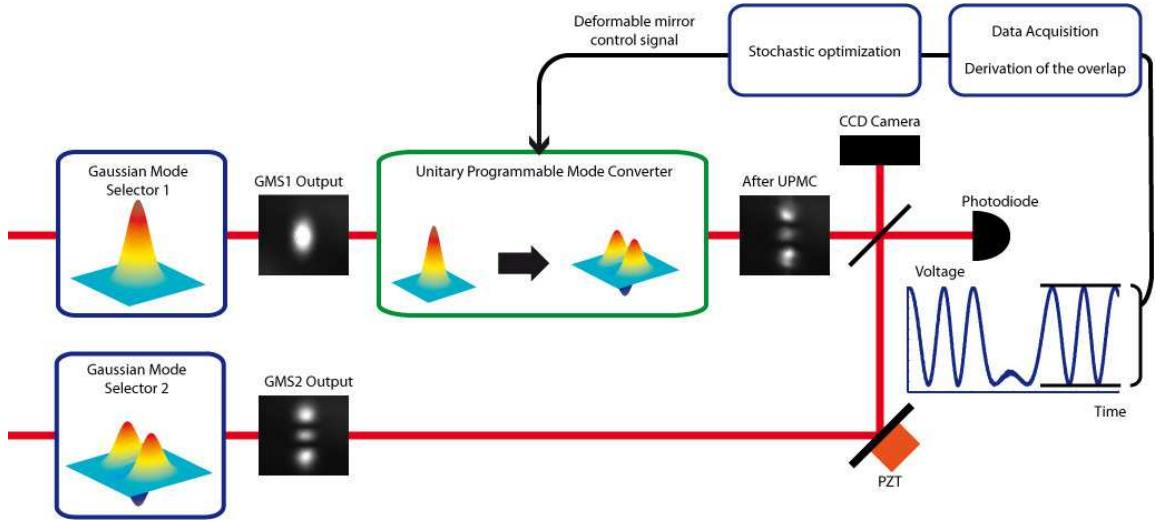}}
\caption{(Color online) The beams coming from the laser go through Gaussian Mode Selectors locked to the desired input and output modes (here $I=TEM_{00}$ and $O=TEM_{20}$).  The desired output is phase modulated using the electroactuator PZT and overlapped with the output of the UPMC. The interference signal is then measured on a photodiode, and the intensity overlap is derived. Depending on the measured overlap, the stochastic optimization algorithm changes the control signal of the deformable mirror.}
\label{GeneralSche}
\end{center}
\end{figure}

Assessing how well the UPMC performs the desired single mode unitary transform requires the stable production of both the input mode (to send into the optical system responsible for the transform) and of the desired output mode (in order to measure the strength of its overlap with the output of the optical system). The mode conversion efficiency, $\alpha^2$, is found through the intensity overlap measurements. The same measurement could be achieved through an intensity and phase profile detection coupled with a computed scalar product, but without the stability and the precision provided by direct experimental measurement of the overlap.\\
We produce stable input and output profiles using mode cleaning cavities operating as gaussian mode selectors, locked to the desired resonating modes: $I(x,y)=TEM_{m0}(x,y)$ and $O(x,y)=TEM_{n0}(x,y)$. Phase plates are placed before the cavities to couple light from the $TEM_{00}$ mode produced by the laser into the desired mode. The cavity is then locked to this mode. This technique is very lossy; a 36\% loss in power results when a $TEM_{00}$ mode is coupled into a cavity locked to the $TEM_{10}$ mode.\\
The output mode of the UPMC is then made to overlap the stable output mode of the reference mode cleaning cavity. The powers of the two outputs are balanced and a mirror mounted on a piezo electric transducer modulates the overall phase of the reference beam, thus providing an interference signal. The visibility of the interference signal is then derived; it is the intensity overlap between the two profiles. This process is described in Fig. \ref{GeneralSche}.\\

The deformable mirror is controlled through 140 computer-controlled actuators, each moving a different part of the gold membrane. The different gains and offsets for each actuator and the coupling between them make the relation between the computer signals and the actual membrane topography difficult to derive. This is a common problem in adaptive optics, and is often solved using active feedback \cite{Shirai:2002p2337}. This is the method employed here; a stochastic optimisation on the computer signals was performed. Depending on the measuremed value of the intensity overlap, we changed stochastically the computer signals, and measured the overlap again. The best value is kept and the process was repeated as fast as was allowed by the settling time of the deformable mirror. The overlap converges to a maximum value, where the process was then stopped.  \\
Power fluctuations arising from the laser make it redundant to introduce additional randomness in the optimization process. Instead, we adopt an optimization routine which moves one actuator at a time. Each individual actuator is moved in order to maximize the intensity overlap, then the next one is moved. The order of the actuators is random. It is trivial to show that for a single reflection, such a sequential optimization will find the maximum intensity overlap possible.\\

\begin{figure}[htbp]
\begin{center}
\mbox{\includegraphics[width=6.00in]{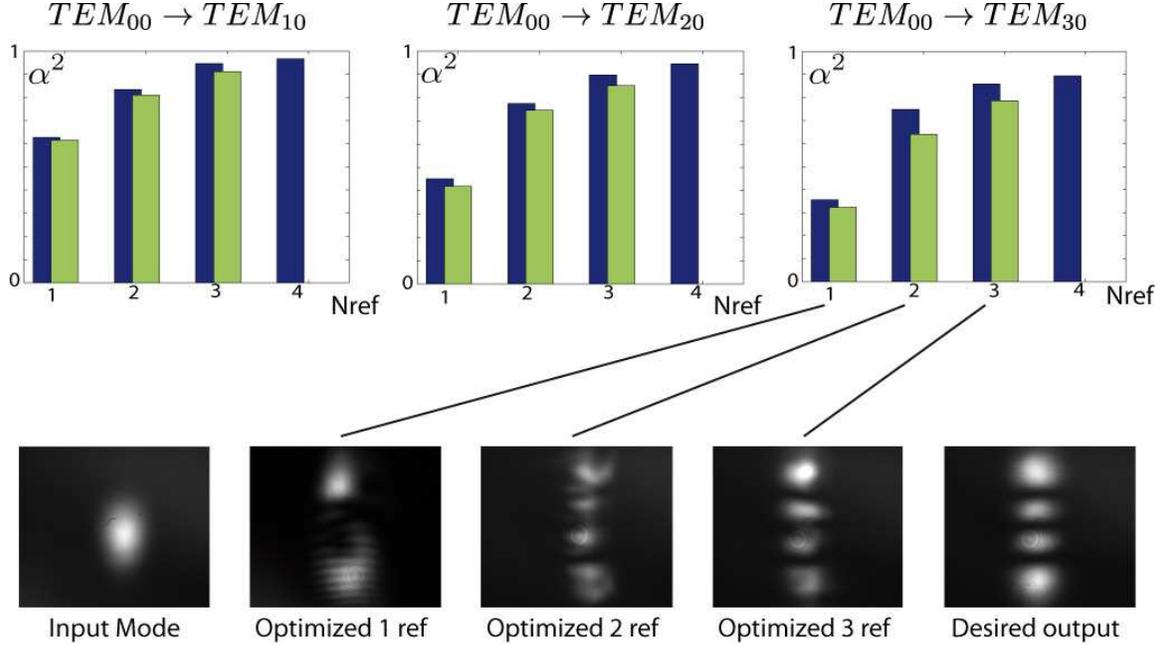}}
\caption{(Color online) Mode conversion efficiency, $\alpha^2$, for three different transformations (green). Simulated results are also plotted (blue). The number of reflections allowed on the UPMC are varied; the screenshots below the plots represent stills from the CCD camera, capturing the output of the UPMC. For repeated optimization procedures of the same transform, with the same number of reflections, the membrane topography was found to differ greatly, while the mode conversion efficiency was consistent. This can be explained by the high number of remaining degrees of freedom. When the maximum mode conversion efficiency was low, i.e. for small number of reflections, the shape of the optimized output mode differed from one optimization to another. In the case of high mode conversion efficiency, most of the power is in the desired output mode, while in the case of low efficiency, a significant portion of the power is not in the desired output mode and changes the intensity distribution of the output depending on the membrane topography.}
\label{Conversions}
\end{center}
\end{figure}

For each desired transform, the gaussian mode selectors are locked to the desired input and output modes. Next, the number of reflections the beam is given on the deformable mirror is controlled by the half wave plates within the UPMC. Now, with the deformable mirror flat, we balance the powers between the output of the UPMC and the reference beam. We then proceed to optimize the membrane topography. 
Fig. \ref{Conversions} presents the mode conversion efficiencies obtained after optimization for different transforms and for different numbers of reflections. As can be clearly seen in the figure, for all the transforms considered, the quality of the conversion consistently improves with the number of reflections allowed. This is in agreement with the fundamental idea underlying the UPMC that successive reflections on a deformable surface eventually lead to a perfect unitary transform. \\
In the case of the single mode transform, the value $\alpha^2$ is the fraction of the power of the output mode effectively converted into the desired mode. For example, the value $\alpha^2$ of $0.91$ measured for the conversion $TEM_{00} \rightarrow TEM_{10}$ with three reflections on the deformable mirror means that $91\%$ of the power of the UPMC's output is in the desired $TEM_{10}$. The largest sources of losses in the UPMC are due to the aforementioned gold coating and the protective window for these three reflections, giving us a loss $12\%$ in optical power. Thus, the limiting factor of the overall efficiency of the physical system in transferring the light from the $TEM_{00}$ input into the $TEM_{10}$ output is the quality of the coatings and surfaces on the optical path, for which there are technical solutions. Reverse transforms were also tested and for the transform $TEM_{10} \rightarrow TEM_{00}$ with 3 reflections the mode conversion efficiency is also $0.91$. \\
The mode conversion efficiency for the transform $TEM_{00} \rightarrow TEM_{50}$ is $0.60 \pm 0.01$; the quality of this transform is limited by the size of the beam on the deformable mirror. The optical set-up was chosen so that the energy of a $TEM_{00}$ is spread over five actuators (see Fig. \ref{ScheUPMC}), therefore a significant portion of the third reflection in the $TEM_{00} \rightarrow TEM_{50}$ transform (that tends to be the size of a $TEM_{50}$) hits the deformable mirror outside of the controllable membrane. This underlines the geometrical limit of this specific UPMC; it can only efficiently handle modes from $TEM_{00}$ to $TEM_{40}$. Since the spot size of a $TEM_{n0}$ mode scales as $\sqrt{n+1}$, an increased number of actuators would allow for the manipulation of more modes with the same precision. \\
Importantly, all of these results were obtained using the same optical set-up. To change from one conversion to another, we simply sent a different computer signal to the deformable mirror. Once an optimal signal is found using the stochastic optimization, the deformable mirror can be repeatably returned to this optimal setting in 10ms. \\

\section{Multimode generalization}

In order to check the capabilities of a realistic UPMC for multimode transforms, we simulate the proof of principle UPMC using a direct propagation model and check the model's validity against the experimental results. We can then use the model to assess this UPMC's performances in manipulating multiple modes and performing general unitary transforms.
Fig. \ref{Conversions} provides a comparison between the simulated and experimental UPMC. The good agreement between the two methods validate the model as a tool to explore the multimode capabilities of this experimental set-up. The limitations on the optimization speed of the experimental set-up compared to the computational simulations explain the small systematic difference between the results: a typical experimental optimization time allows for $10^5$ trials. Computational tests involve typically $10^6$ to $10^7$ trials. By simulating the experimental optimization, we found that the $10^2$ ratio explains the systematic difference.\\

We simulate the UPMC by modelling the propagation of the light field. The light profile is input as a 2048 value array. The FT is performed as a normalized and centered Fast Fourier Transform aglorithm. The reflection on the deformable mirror is a product element per element with a 2048-element array of phases. This array of phases derives from the 12 phase values of the actuators, with smoothing between them to take into account the continuous nature of the membrane. 
We use a stochastic approach (simulated annealing) to find the optimal phase profile for a specified transform.\\
The simulation performed was on a practical optical system for single and two mode conversions for a wide set of modes. For all these transforms we considered from one to four reflections on the deformable mirror, separated by FTs. We mainly considered transforms between $TEM_{mn}$ modes, with the exception being a flip mode (a $TEM_{00}$ with a $\pi$ phase shift in the middle).\\

\begin{figure}[htbp]
\begin{center}
\mbox{\includegraphics[width=4.00in]{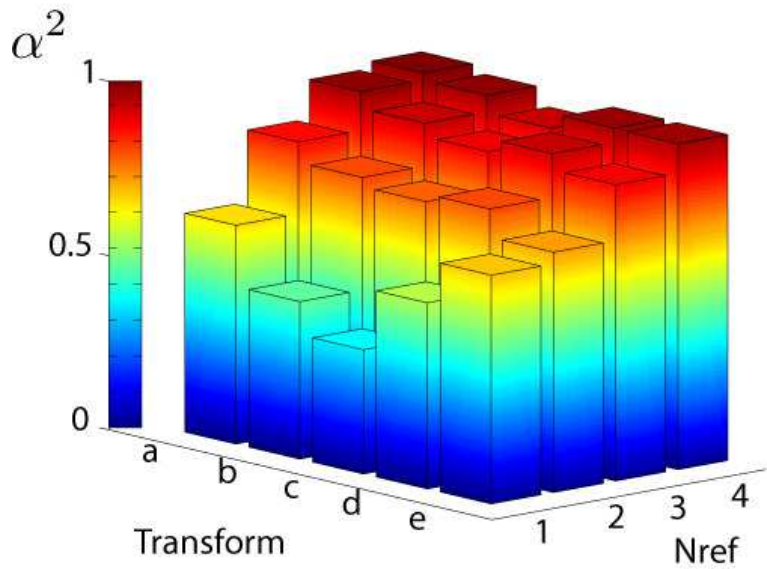}}
\caption{(Color online) Presentation of the mode conversion efficiency $\alpha^2$ as a function of the transform considered and the number of reflections Nr allowed. The transforms are a. $TEM_{00} \rightarrow TEM_{10}$ b. $TEM_{00}  \rightarrow TEM_{20}$ c. $TEM_{00}  \rightarrow TEM_{30}$ d. $TEM_{10}  \rightarrow TEM_{30}$ e. $TEM_{10}$ to flip mode}
\label{SingGaussian}
\end{center}
\end{figure}

\begin{figure}[htbp]
\begin{center}
\mbox{\includegraphics[width=4.00in]{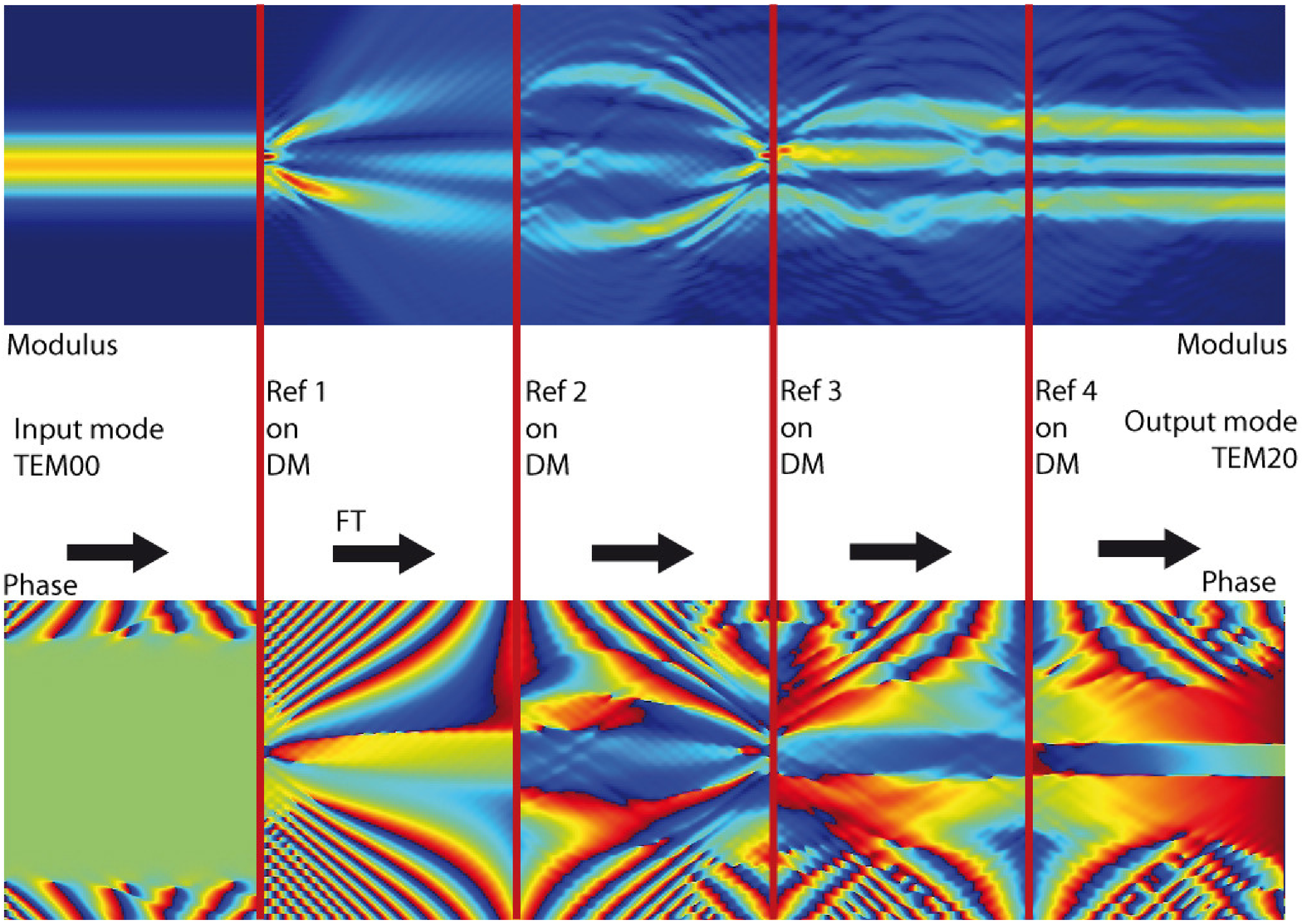}}
\caption{(Color online) Transverse profiles of the magnitude of the field when undergoing a succession of FTs and reflections on deformable mirrors (DM). The magnitude remains constant at the reflection surfaces while the phase has a sharp discontinuity. The transverse axis is renormalized to keep the profile of $TEM_{00}$ constant throughout propagation.}
\label{NiceLookingModeConversion}
\end{center}
\end{figure}

Fig. \ref{SingGaussian} presents simulated mode conversion efficiencies for single mode transforms. Among them a., b.  and c. are used in Fig. \ref{Conversions} to check the validity of the simulations. The efficiency consistently increases with the number of reflections allowed, for all the transforms considered. The results for single reflections match the theoretical maximums (modulus overlap). Comparing the results of d. and e. to c. and a. respectively show that the difference between the shape of the input mode and the shape of the desired output has a stronger impact on the overlap than the complexity of the modes themselves. 
Fig. \ref{NiceLookingModeConversion} presents the magnitude and phase evolution in the conversion process: $TEM_{00} \rightarrow TEM_{20}$. This simulation details the spatial process that the light undergoes. On each surface, the intensity profile is not changed, but a phase is printed onto the light field. The propagation of the phase profile is then responsible for the change in the intensity profile of the beam.\\

The propagation model is now generalized to simulate multimode manipulations by the UPMC. For a desired transform D, we optimize the topography of the deformable mirror to make the UPMC perform D. The transform quality $\alpha_n^2$ evaluate how closely the UPMC's transform matches the desired transform. To compute $\alpha_n^2$, the multiple input modes are propagated throuh the simulated UPMC sequentially. For each of the output modes, the intensity overlap with the desired output mode is computed. The overlaps are then combined to form the transform quality $\alpha_n^2$. The simulated annealing process optimizes the topography of the deformable mirror so as to improve $\alpha_n^2$, until a maximum value is reached.  \\

Multimode manipulations can take many forms. We focus here on two families of transforms: phase operators; and beamsplitters. Two mode transforms between the first two TEM modes will be considered - $TEM_{00}$ and $TEM_{10}$. The beamsplitter transform is the matrix 
\begin{equation}
U_{BS}(r)=
\begin{pmatrix}
r & t \\
t & -r \\
\end{pmatrix}
\end{equation}
with the relationship $t=\sqrt{1-r^2}$. $U_{BS}(r)$ is the beamsplitter matrix for a half silvered mirror of reflectivity $r^2$.
The phase operator is defined by the matrix
\begin{equation}
U_P(\phi)=
\begin{pmatrix}
1 & 0 \\
0 & e^{i\phi} \\
\end{pmatrix}
\end{equation}
and corresponds to introducing a phase shift between the two copropagating modes.\\

\begin{figure}[htbp]
\begin{center}
\mbox{\includegraphics[width=6.00in]{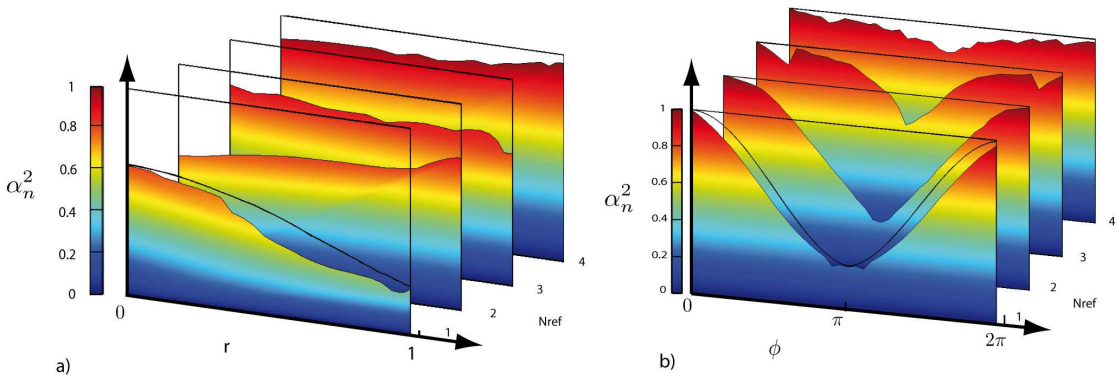}}
\caption{(Color online) Transform quality $\alpha_n^2$ when the realistic UPMC is optimized to perform a beamsplitter $U_{BS}$ (a) and a phase operator $U_P$ (b). They are plotted as a function of the number of reflections allowed and their respective parameters $r$ and $\phi$. The black curve represents the single reflection theoretical maximum. The overlap is perfect for $\alpha_n^2=1$. Additional simulations with the same number of trials for a constant transform $D$ presented the same small fluctuations in the case of three and four reflections. These are artefacts of the optimization process.}
\label{Twomode}
\end{center}
\end{figure}

Fig. \ref{Twomode} presents the best transform quality achieved, $\alpha_n^2$, when the UPMC is optimized to perform the transforms  $U_{BS}(r)$ and $U_P(\phi)$.
We present $\alpha_n^2$ as a function of $r$ and $\phi$ and the number of reflections allowed. In the single reflection case, we can derive the theoretical maximum for the transform quality, assuming the topography is fully controllable (i.e. not limited to a twelve actuators control). Comparing the performances of the realistic UPMC to the theoretical maximums shows that the limited number of actuators has an impact on the transform quality. On the other hand, when higher number of reflections are allowed, the UPMC outperforms the single reflection theoretical maximums. With increased number of reflections, the efficiency improves for all values of $r$ and $\phi$: this validates the multiple reflection scheme as a way to perform unitary transforms. \\

The high $\alpha_n^2$ values obtained both for the beamsplitters and the phase operators for 3 and 4 reflections, especially compared to the single mode transforms in Fig. \ref{SingGaussian} tends to show that the limiting factor is the complexity of the modes manipulated, rather than the number of manipulated modes. This means that this realistic UPMC can efficiently manipulate multiple copropagating modes, mixing them or introducing phase shifts between them.\\

As the constraints include more modes, the transform quality gets lower, but the tendancy remains: with more reflections allowed, the unitary transform performed by the UPMC approaches the desired one. This statement is in agreement with the theoretical result, and shows empirically that for a limited number of resources (i.e. reflections and actuators), efficient transforms are possible. \\

\section{Conclusion}

As a conclusion, we showed that multiple reflections on deformable mirrors separated by FTs can perform any unitary spatial transformation on a beam of light. Realistically,  the capabilities of this unitary programmable mode converter is only limited by the number of reflections allowed and the number of pixels for each reflection. We show experimentally that three reflections on an array of 12 pixels are enough to perform single mode transforms with an efficiency better than $80\%$ for the first 4 modes of the TEM basis. This achievement also validates our model for the UPMC, allowing us to compute the efficiency of multimode transforms, the simultaneous transformation of multiple input modes. We find that the UPMC can perform multiple mode manipulations as well as single mode manipulations. 
The programmable nature of the UPMC makes it a good candidate for general light manipulation. An important application is its uses after a multi-mode fiber as a way to compensate mode diffusion. Moreover, since all the losses in the UPMC can be reduced by technical improvement of the quality of the coatings and surfaces, it opens general spatial light manipulation to quantum protocols. \\

\section{Acknowledgements}
This work was funded by the Centre of Excellence program of the Australian Research Council. It was supported by the Australian Research Council Discovery Project DP0985078 and the NCI National Facility at the ANU. We acknowledge the financial support of the Future and Emerging Technologies (FET) programme within the Seventh Framework Programme for Research of the European Commission, under the FET-Open grant agreement HIDEAS, number FP7-ICT-221906.

\newpage
\bibliographystyle{osajnl}

\end{document}